\documentclass[apjl]{emulateapj}
 \slugcomment{{\sc Accepted to ApJ Letters:} March 17, 2014}
\usepackage[utf8]{inputenc}
\usepackage{natbib} 
\usepackage{soul} 

\begin{document}

\title{
Five Debris Disks Newly Revealed in Scattered Light from the HST NICMOS Archive}
\author{R\'emi Soummer\altaffilmark{1}, Marshall D. Perrin\altaffilmark{1}, Laurent Pueyo\altaffilmark{1,4}, \'Elodie Choquet\altaffilmark{1}, Christine Chen\altaffilmark{1}, David A. Golimowski\altaffilmark{1}, J. Brendan Hagan\altaffilmark{1,3}, Tushar Mittal\altaffilmark{2,4},  Margaret Moerchen\altaffilmark{1}, Mamadou N'Diaye\altaffilmark{1}, Abhijith Rajan\altaffilmark{6}, Schuyler Wolff\altaffilmark{1,3}, John Debes\altaffilmark{1}, Dean C. Hines\altaffilmark{1}, Glenn Schneider\altaffilmark{5}}
\altaffiltext{1}{Space Telescope Science Institute, 3700 San Martin Dr, Baltimore MD 21218}
\altaffiltext{2}{Now at UC Berkeley, Berkeley CA 94720}
\altaffiltext{3}{Now at Purdue University, West Lafayette IN, 47907}
\altaffiltext{4}{Johns Hopkins University, 3400 North Charles Street, Baltimore MD 21218}
\altaffiltext{5}{Steward Observatory, The University of Arizona, 933 North Cherry Avenue, Tucson, AZ 85721}
\altaffiltext{6}{Arizona State University, Phoenix, AZ 85004}

\begin{abstract}
We have spatially resolved five debris disks (HD~30447, HD~35841, HD~141943, HD~191089, and HD~202917) for the first time in near-infrared scattered light by reanalyzing archival \emph{Hubble Space Telescope} (HST)/NICMOS coronagraphic images obtained between 1999 and 2006. One of these disks (HD~202917) was previously resolved at visible wavelengths using HST/Advanced Camera for Surveys.  To obtain these new disk images, we performed advanced point-spread function subtraction based on the Karhunen-Lo\`eve Image Projection (KLIP) algorithm on recently reprocessed NICMOS data with improved detector artifact removal (Legacy Archive PSF Library And Circumstellar Environments Legacy program). Three of the disks (HD~30447, HD~35841, and HD~141943) appear edge-on, while the other two (HD~191089 and HD~202917) appear inclined. The inclined disks have been sculpted into rings; in particular, the disk around HD~202917 exhibits strong asymmetries.  All five host stars are young (8--40 Myr), nearby (40--100 pc) F and G stars, and one (HD~141943) is a close analog to the young sun during the epoch of terrestrial planet formation.  Our discoveries increase the number of debris disks resolved in scattered light from 19 to 23 (a 21\% increase).  Given their youth, proximity, and brightness ($V = 7.2$ to 8.5), these targets are excellent candidates for follow-up investigations of planet formation at visible wavelengths using the \emph{HST}/STIS coronagraph, at near-infrared wavelengths with the Gemini Planet Imager (GPI) and Very Large Telescope (VLT)/SPHERE, and at thermal infrared wavelengths with the \emph{James Webb Space Telescope} NIRCam and MIRI coronagraphs.

\end{abstract}


\section{Introduction}

Infrared surveys have identified more than a thousand nearby star systems for which infrared excesses beyond $\sim 10~\mu$m reveal circumstellar dust produced from the collisional grinding and destruction of small planetesimals. The amount of dust generally decreases with age because small grains are removed by radiation pressure and Poynting-Robertson drag.  Without gas to retard their removal, circumstellar grains typically possess lifetimes of 10,000 yr, which is significantly shorter than the age of the star.   These circumstances suggest that the grains are replenished from a reservoir of unseen planetesimals such as asteroids or comets, which are perturbed into orbits that lead to dust-generating collisions.  Coronagraphic imaging of debris disks has revealed structures such as localized brightness peaks, asymmetries, and warps that suggest the presence of formed or forming planets \citep{Wyatt:2008p2687}. The first three systems with directly imaged exoplanets (Fomalhaut, HR~8799, and $\beta$~Pictoris; \citealp{2008Sci...322.1345K,Marois:2008p2921,Lagrange:2010p3211}) are all stars previously known to host debris disks.
For instance, recent ground-based, high-contrast imaging has discovered a $9 \pm 3~M_{Jup}$ planet orbiting 8--13~AU from $\beta$~Pictoris that is consistent with a warp or secondary disk observed in scattered light \citep{Golimowski:2006p501,Lagrange:2010p3211}.

Modeling the dynamical effects inferred from scattered-light morphologies places constraints on the architectures of exoplanetary systems \citep{2006ApJ...648..652S}.   Models based solely on spectral-energy distributions (SEDs) are inherently degenerate, so direct images are essential to locate the dust and planetesimal belts (the analogs of our asteroid and Kuiper belts) unambiguously.  Furthermore, models based only on assumptions about grain sizes and compositions yield disk radii and dust masses that may vary by an order of magnitude \citep{2006ApJ...638.1070H}.  However, multi-band scattered-light images of the disks provide measurements of the colors, phase functions, and albedos of the dust that can be used by the models to constrain the physical properties of the dust grains \citep{Golimowski:2006p501,2007ApJ...654..595G,2009ApJ...696.2126S}.

Only 23 debris disks have published spatially resolved detections in scattered light as of early 2014, primarily using coronagraphs aboard the \emph{Hubble Space Telescope (HST)} \citep[e.g.,][]{Weinberger:1999p175,Schneider:1999p532,Clampin:2003p118,Ardila:2004p451,Krist:2005p511,Golimowski:2006p501,SSH06,Krist07,Krist:2010p3152,2011AJ....142...30G}. This subset of known debris disks comprises 
the 16 disks listed in Table 1 of \citet{2011AJ....142...30G}, HD 202628 \citep{2012AJ....144...45K}, HIP 79977 \citep{2013ApJ...763L..29T}, and the five disks reported in this {\it Letter}. Scattered-light images typically possess higher angular resolution than images in thermal emission, and their increased sensitivity to micron-sized particles at large distances from the star provides more detailed information about the spatial distribution of the smallest dust grains.  Together, scattered-light and thermal-emission images can be used to constrain the azimuthal dependence of the dust density distribution and the properties of the constituent grains (e.g., composition, size, porosity).

The principal difficulty with obtaining such images of disks or planets -- even with the highly stable \emph{HST} coronagraphs or the latest generation of specialized ground-based instruments (GPI, SPHERE, Project 1640, HiCIAO) \citep{2008SPIE.7015E..31M,2008SPIE.7014E..41B,2008SPIE.7014E..42H,2011PASP..123...74H} or the \emph{James Webb Space Telescope (JWST)} NIRCam and MIRI coronagraphs \citep{2010PASP..122..162B} -- is the large contrast between the faint scattered light from the disk and the much brighter halo of starlight from the instrumental point-spread function (PSF). The residual starlight must be precisely calibrated and subtracted during image processing.  

Recent advances in coronagraphic image processing have been made through the development of sophisticated algorithms for removing the residual diffracted light, which utilize large libraries of reference coronagraphic stellar PSFs \citep[e.g., ][]{Lafreniere:2007p274,LMD09}). 

Recently, we initiated the Archival Legacy Investigation of Circumstellar Environments (ALICE) project to reprocess comprehensively and consistently archived images from various \emph{HST}/NICMOS coronagraphic surveys for
faint circumstellar companions. This program uses the Karhunen-Lo\`eve Image Projection (KLIP) algorithm \citep{Soummer:2012p3139} and the large number of reference stars available in the \emph{HST} archive.
KLIP not only decreases computational costs incurred from previous optimization methods, but also improves the ability to detect circumstellar disks by greatly mitigating the disk self-subtraction problem that plagued earlier algorithms.

\begin{figure}[htbp]
\center
\resizebox{\hsize}{!}{\includegraphics{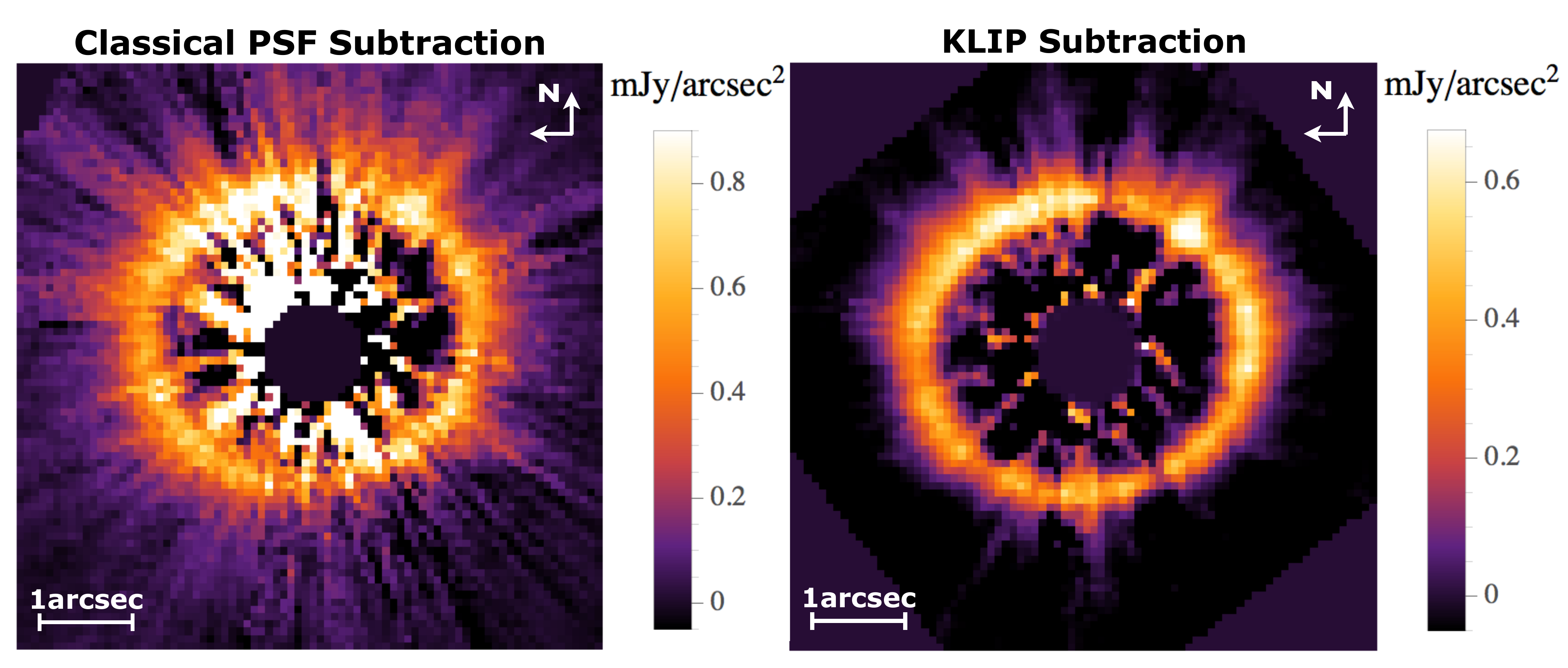}}
 \caption{Improved residual starlight subtraction using the KLIP algorithm in the ALICE pipeline for the well-known debris disk around HD~181327. The left image was produced using conventional subtraction techniques \citep{SSH06}, while the right image is obtained using KLIP \citep{Soummer:2012p3139} and a library of reference PSFs from the LAPLACE Archive \citep{SSS10}. KLIP significantly improves the  subtraction within 1\farcs5 of the star, as evidenced by the lower residuals within the HD~181327 ring. The newly detected disks presented in this paper lie within this range of angular separation, which explains why they were not seen previously in the NICMOS images. At an angular separation of 1 arcsec, KLIP improves the coronagraphic image contrast by a factor of $\sim50$ over classical PSF subtraction, based on average results from over 7 different coronagraphic images of stars without known circumstellar disks. The reduced apparent surface brightness with KLIP (algorithm throughput) can be calibrated with forward modeling \citep{Soummer:2012p3139}. }\label{Fig:KLIP}
\end{figure}

Our ALICE pipeline improves the coronagraphic detection limit for point sources by at least an order of magnitude over classical PSF subtraction methods.  For disk images, most of the improvement is obtained at small separations, as demonstrated in Figure \ref{Fig:KLIP} for the previously imaged debris disk
around HD~181327 \citep{SSH06}.

In this \emph{Letter}, we report some of the first results from ALICE: spatially resolved images of five debris disks that were previously undetected in archived NICMOS data. Four of these disks have never before been imaged in scattered light. The disk around HD~202917 was previously detected by the \emph{HST} Advanced Camera for Surveys (ACS) Guaranteed Time Observer team and reported by \citet{Krist07} at the 2007 Spirit of Lyot conference. The images and analyses presented here constitute  ``first looks" at these systems.  Each image warrants more in depth analysis, and detailed follow-up papers are in preparation.

\section{Data and Data Processing}

\subsection{Archival Data Sets}

Between \emph{HST} Cycles 7 and 15, a total of 215 \emph{HST} orbits were devoted to six NICMOS surveys and searches for debris disks. 
The five star--disk systems newly resolved by ALICE (Figure \ref{Fig:5disks}) were identified from data obtained by these programs (Table 1).   Although the principal goal of the ALICE project is detecting faint companions in these NICMOS data, the advanced starlight subtraction enabled by the KLIP algorithm has allowed us to also detect extended scattered light from disks with unprecedented sensitivity.

\begin{deluxetable*}{l|cccc|cc|ccc|ccc}
\tabletypesize{\footnotesize}
\tablecaption{System properties and \emph{HST} program information\label{SummTable}}
\tablecolumns{13}
\tablewidth{0pc} 
\tablehead{
 & & & & \colhead{$L_{IR}/L_\star$\tablenotemark{a}} & \colhead{Dist.} & & & \colhead{Age} & & \colhead{\emph{HST}} & & \colhead{Observation}\\
\colhead{Name} & \colhead{SpT} & \colhead{$V$} & \colhead{$B$--$V$} & \colhead{($\times10^{-4}$)} & \colhead{(pc)} & \colhead{Ref.} & \colhead{Association} & \colhead{(Myr)} & \colhead{Ref.} & \colhead{Program} & \colhead{Filter} & \colhead{Date}
}
\startdata
HD~30447 & F3V & 7.86 & 0.39 & ~~~7.9 & $80 \pm 5$ & 1 & Columba & 10--40 & 5,6 & 10177 & F110W & 2005  \\
HD~35841 & F3V & 8.90 & 0.41 & 13 & 96 & 2 & Columba & 10--40 & 5,6 & 10177 & F110W & 2005  \\
HD~141943 & G2V & 7.85 & 0.63 & ~~~1.2 & $67 \pm 7$ & 3,4 & \nodata & 17-32 & 4,8,9 & 10176,10527  & F160W,F110W & 2006,2007 \\
HD~191089 & F5V & 7.18 & 0.44 & 13 & $52 \pm 1$ & 1 & $\beta$~Pic & 8--20 & 7 & 10527  & F110W & 2007 \\
HD~202917 & G7V & 8.67 & 0.65 & ~~~2.5 & $43 \pm 2$ & 1 & Tuc--Hor & 10--40 & 5,6 & 7226,10849 & F160W,F110W & 1999,2007
\enddata

\tablenotetext{a}{Ratio of the disk's integrated infrared luminosity and the star's bolometric luminosity derived from the SEDs shown in Figure \ref{Fig:SEDs}. \newline
REFERENCES.-- (1) \citealt{1997yCat.1239....0E}; (2) \citealt{2011ApJS..193....4M}; (3) \citealt{2005ASP...338..280M}; (4) \citealt{2008ApJ...677..630H}; (5) \citealt{2006ApJ...644..525M}; (6) \citealt{2013ApJ...762...88M};  (7) \citealt{2004ApJ...603..738Z}; (8) \citealt{2011MNRAS.413.1922M}; (9) \citealt{2011MNRAS.413.1939M}};
\end{deluxetable*}

\subsection{Data Reduction}
Conventional PSF subtraction techniques employ contemporaneous images of a reference star whose colors and brightness closely match the target star and/or contemporaneous images of the target star itself obtained after rolling the telescope's field of view around the coronagraphic axis \citep{Schneider:1999p532}.  Starlight subtraction can be improved considerably by using a linear combination of many reference images to synthesize an optimal reference PSF \citep[e.g., LOCI;][]{Lafreniere:2007p274}.  A library of references can be assembled either by observing a large number of reference stars or by introducing observational diversity, such as angular or spectral differential imaging \citep{MArois2006,MMV10}.  Subsequent variants of LOCI have modified the optimization parameters for better performance in these various regimes \citep{Thalmann:2010p3175,SHP11,Pueyo:2012p3205}.  More recently, new optimization techniques based on principal component analysis (PCA), such as the KLIP algorithm used in the ALICE pipeline, have been proposed \citep{Soummer:2012p3139,2012MNRAS.427..948A,OBB13}.  The use of such advanced optimization techniques is particularly well suited to \emph{HST} archival research, as a large library of reference stars already exists.

The \emph{HST} archival program entitled ``Legacy Archive PSF Library And Circumstellar Environments'' (LAPLACE) recently recalibrated the entire NICMOS coronagraphic archive from Cycle 7 through Cycle 15 \citep{SSS10}. LAPLACE improved on prior calibrations by using contemporary flat-field frames and observed dark frames (instead of epochal flats and synthetic darks) and by implementing better bad-pixel correction.  LAPLACE also provided important ancillary information about each target star (e.g., its position behind the coronagraphic mask and its $JHK$ magnitudes) that facilitates the selection and normalization of reference images.  

We use the KLIP algorithm to decompose the LAPLACE reference PSF library into an orthogonal basis of eigen-images via a Karhunen-Lo\`eve transform and then truncate that basis to retain the $n$ modes with the highest weights.  A synthetic reference PSF model is computed from vector projections of the actual science images into the truncated PCA basis.  The modeled PSFs are then subtracted from the science images.  

The ALICE pipeline provides tools for selecting optimal reference star subsets and number of modes $n$ for each subtraction. In addition to KLIP processing, ALICE uses iterative roll subtraction \citep{2010AJ....140.1051K} to enhance the disk detection when combining data from multiple spacecraft orients.

\begin{figure*}[htbp]
\center
\includegraphics[width=6in]{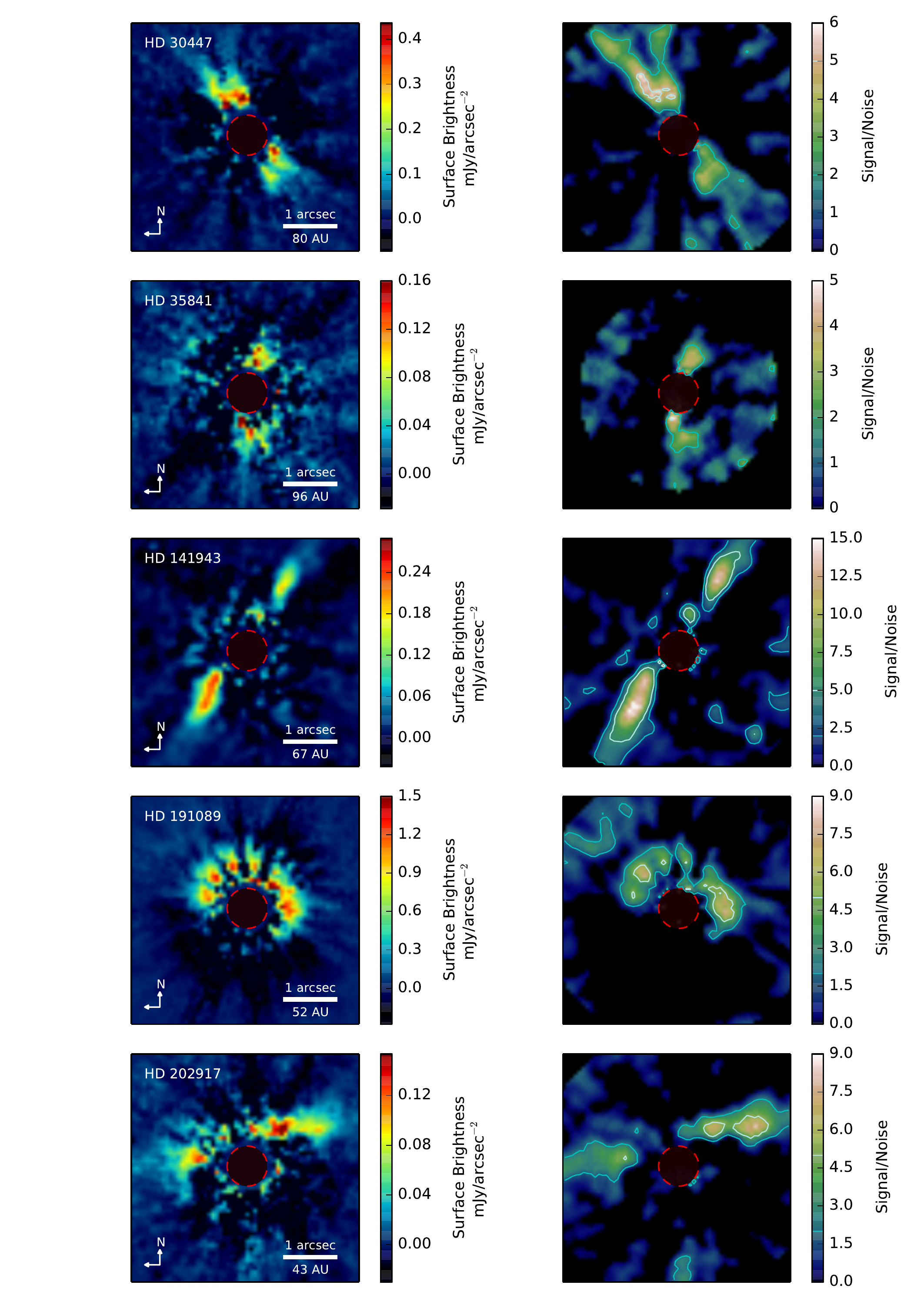}
\caption{Scattered-light images in F110W (1.1 micron) of five debris disks found in the NICMOS coronagraphic archive after processing with KLIP. The surface brightnesses shown in mJy/arcsec$^2$ correspond to \emph{lower limit} estimates mainly because of algorithm throughput. The first three disks appear nearly edge-on while the latter two appear to be rings at lower inclinations.  HD~202917's disk was previously resolved in \emph{HST}/ACS coronagraphic images. Noise was estimated based on an ensemble of non-detections as described in Section 2.2. The provided SNR maps are calculated for an aperture of two resolution elements in diameter to smooth slightly the local variations. }\label{Fig:5disks}
\end{figure*}

Figure \ref{Fig:5disks} shows the NICMOS images of the five stars listed in Table~1 processed through the ALICE pipeline that reveal scattered-light disks extending to angular distances of $1''$--$2''$ from each star.   All five disks are seen in images obtained at two \emph{HST} roll orientations, and the two disks (HD~141943 and HD~202917) for which multi-band archival data is available are seen in both bandpasses (F110W and F160W).  Only F110W images are shown in Figure \ref{Fig:5disks} for consistency between all targets and because the F110W images have better angular resolution and sensitivity. 

Also shown are maps of the signal-to-noise ratio (SNR). The noise is calculated from a representative ensemble of reference stars with non-detections that were processed consistently with the science targets (i.e. subtracted using the same number of PCA modes, rotated, combined). These references are used to build a pixel-to-pixel noise covariance matrix and estimate the noise variance at every position within the same aperture as used for signal extraction (two resolution elements in diameter). These are the same methods as being implemented for JWST coronagraphy (Pontoppidan et al. in prep; Pueyo et al. in prep). The SNR displayed in Figure 2 illustrates the statistical confidence of our disk detections with respect to the larger ensemble of sources processed by the ALICE program. This approach rigorously accounts for both temporal variations and spatial correlations (e.g. speckles or diffraction spike subtraction residuals). While the SNR in any individual location is modest in these data, the integrated SNR over each disk image is statistically robust. 

 The surface brightnesses shown in Figure \ref{Fig:5disks} are lower limits due to the throughput and zero-mean output of the KLIP algorithm.  Indeed the disk itself can contribute non-negligibly to the mean image.  These effects can be well calibrated using forward modeling \citep{Soummer:2012p3139}, but such modeling is beyond the scope of this paper and is deferred to separate detailed analyses of the disks in subsequent papers. 
In any case these surface brightness estimates should be close within less than a factor 2, based on the example shown in Figure 1 where the KLIP image surface brightness is $\sim 75\%$ of the classical PSF subtraction image. See also Figure 2 in \citet{Soummer:2012p3139} for an estimate of the algorithm throughput.

\begin{figure}[htbp]
\center
\resizebox{\hsize}{!}{\includegraphics{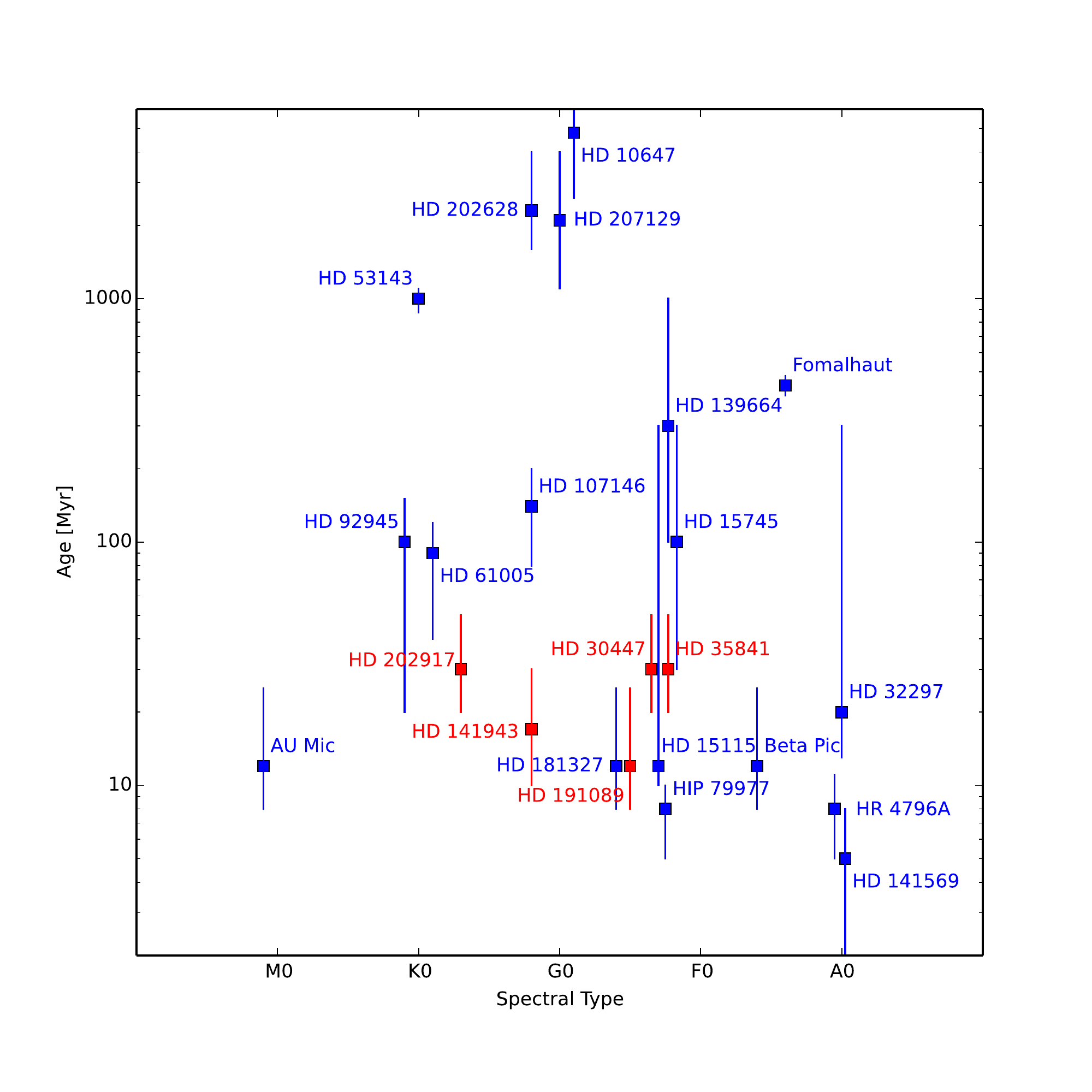}}
 \caption{Ages and spectral types of the 23 main-sequence stars with debris disks resolved in scattered light.  The five stars hosting the disks presented here (marked in red) are young ($<40$ Myr) and have near-solar spectral types (G2-F3) because of sample biases in the NICMOS surveys from which the images originated.}\label{Fig:agetype}
\end{figure}

\section{Preliminary Analyses of Disk Properties}

\begin{figure*}
\center
\resizebox{\hsize}{!}{\includegraphics{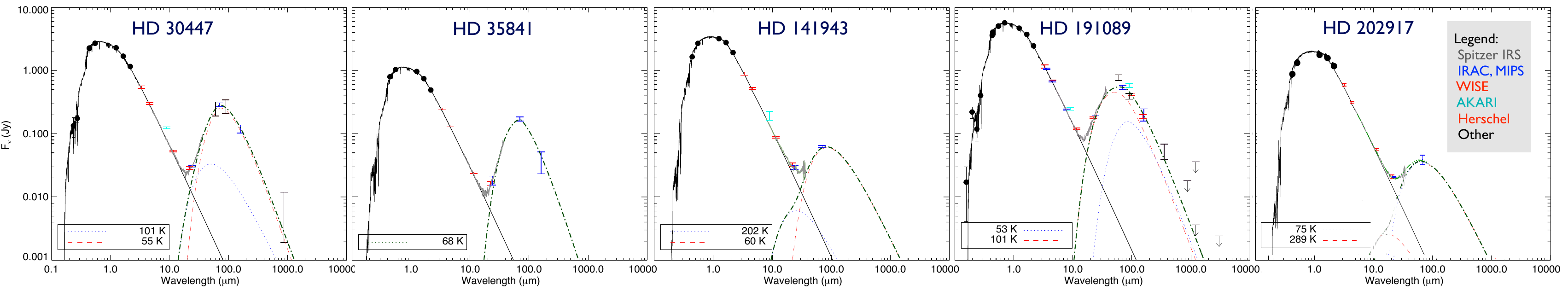}}
\caption{ Spectral energy distributions (SEDs) of the debris-disk systems resolved by ALICE.  The photometric data are plotted in colors matching the sources identified in the legend.  The solid curves represent the model photospheres of \citet{1993yCat.6039....0K} that best fit the stellar photometry. The dashed curves represent one- or two-component black body models that were fitted to the excess infrared emission.  The black body curves provide estimates of the dust temperature assuming radiative equilibrium.  Bayesian evidence criteria indicate that two-component fits are statistically more significant in all cases except HD~35841. These results suggest the presence of multiple belts of dusty material, similar to the Sun's asteroid and Kuiper Belts.}\label{Fig:SEDs}
\end{figure*}

The five disks surround F and G stars whose ages ($< 40$ Myr) correspond to the epoch of terrestrial planet formation in our solar system.  The systems lie in an important region in spectral-type-versus-age space (Figure \ref{Fig:agetype}), as only three debris disks have been previously imaged in scattered light around F or G stars younger than 100 Myr.  Figure \ref{Fig:SEDs} shows the SEDs of the star--disk systems, which have fractional dust luminosities of $10^{-4} \lesssim L_{IR}/L_\star \lesssim 10^{-3}$ that are typical for young systems.  Blackbody fits to the infrared excesses are consistent with large dust-grain populations orbiting at 20--30~AU with characteristic temperatures of $T_{\rm eff} \approx 50$~K.

HD~30447 is a member of the Columba moving group \citep{2011ApJS..193....4M,2013ApJ...762...88M}, which also contains the exoplanet host star HR 8799.  Its disk appears nearly edge-on with a position angle PA~$\approx 35\pm 5^{\circ}$.  The disk extends between projected distances of $\sim 0\farcs75$--2\farcs5 (60--200~AU) from the star.  The brightness and extent of the northeast side of the disk are about twice those of the southwest side.  The maximum SNR on the northeast side is $\sim6$, with a maximum surface brightness of 0.6~mJy~arcsec$^{-2}$. 
The system has an infrared excess $L_{IR}/L_\star \approx 7.9 \times 10^{-4}$, and its SED is best fitted with a two-component dust model with grain temperatures of 55~K and 101~K.

HD~35841 is also a member of the Columba moving group \citep{2011ApJS..193....4M,2013ApJ...762...88M}.  Its disk is detected within 1\farcs5 (144~AU) of the star. It is the faintest and most compact disk in our sample.  The disk appears nearly edge-on, though the two lobes are not diametrically aligned (PAs~$\approx 180^{\circ}$ and $\approx 335^{\circ}$). 
The maximum SNR is $\sim4$, with a maximum surface brightness of 0.15 mJy/arcsec$^2$. 
The system has an infrared excess $L_{IR}/L_\star \approx 1.3\times10^{-3}$, and its SED is best fitted with a single-component dust model with a grain temperature of 68K.   

HD~141943 is a magnetically active and rapidly rotating star whose age \citep[$\sim 17-32$~Myr;][]{2011MNRAS.413.1922M,2011MNRAS.413.1939M,2008ApJ...677..630H} and G2V spectral type are analogous to the young sun during the epoch of terrestrial planet formation. 
Current age estimates suggest that it is younger than the debris-disk G2V host star HD~107146, though the latter could be as young as 30 Myr \citep{Ardila:2004p451,2004ApJ...604..414W}.
Our preliminary models indicate that HD~141943's disk is seen nearly edge-on (inclination 85$^\circ$) at a position angle (PA) of 145$^\circ$. The disk is detected at projected distances of 0\farcs7--2\farcs5 (47--167~AU) from the star.  The maximum SNR on the southeast side is $\sim15$, with a maximum surface brightness of 0.25~mJy~arcsec$^{-2}$. The disk's morphology and brightness are the most symmetric of the disks in our sample, and it appears to be the flattest (i.e., edge-on and vertically unresolved).   The system has the smallest infrared excess in our sample ($L_{IR}/L_\star \approx 1.2\times10^{-4}$), and its SED is consistent with a planetesimal belt a few times more massive than the Kuiper Belt at an age of $\sim10$ Myr \citep{2008ApJ...687..608K}. The SED is best fitted by two dust components with grain temperatures of 60~K and 202~K.

HD~191089 is a member of the $\beta$~Pic moving group \citep{2006ApJ...644..525M}, making it a sibling of the famous disk and planet host $\beta$~Pic \citep{Lagrange:2010p3211}. It has an infrared excess of $L_{IR}/L_\star \sim 1.3 \times 10^{-3}$, and its SED is best fitted by a two-component dust model with grain temperatures of 53~K and 101~K.  The ring-like disk is moderately inclined by $\sim30^\circ$ to the line of sight, and a significant brightness asymmetry suggests the presence of strongly forward-scattering grains.  The projected major axis of the ring has PA~$\approx 60^{\circ}$, which is consistent with the extended thermal emission observed at 18 $\mu$m by \citet{Churcher:2011p3209}. The disk is detected within 1\farcs4 (73~AU) of the star.   The maximum SNR is $\sim7$, with a maximum surface brightness of $1.3$~mJy/arcsec$^2$. The center of the ring appears offset from the star by a few tenths of an arcsecond, a marginal feature that will be explored further with our upcoming HST STIS observations.

HD~202917 is a member of the Tucana-Horologium association with an infrared excess $L_{IR}/L_\star \sim 2.5 \times 10^{-3}$.   Its SED is best fitted with a two-component dust model with grain temperatures of 75~K and 289~K.  The debris disk was previously resolved at visible wavelengths in formally unpublished images obtained with \emph{HST's} ACS coronagraph \citep{Krist07}.  Our reprocessed NICMOS images taken through two filters and two telescope orientations robustly confirm the preliminary ACS detection, including the strong asymmetry between the east and west sides. The F110W detection extends to projected distances approximately 2\farcs5 (107~AU) from the star.  The disk exhibits an asymmetric arc suggestive of a partial ring inclined $\sim 70^\circ$ to the line of sight with a major axis PA~$\approx 300^{\circ}$. The northwest side is significantly brighter and more extended than the southeast side, which together with the similarly asymmetric ACS image suggests a highly perturbed disk. The maximum SNR is $\sim8$, with a maximum surface brightness of 0.2 mJy/arcsec$^2$. 

\section{Concluding Remarks}
We have obtained new scattered-light images of five debris disks found in the HST NICMOS coronagraphic archive after reprocessing those data with the KLIP algorithm in our ALICE pipeline.  Preliminary descriptions of the disks' characteristics are given based on scattered-light image morphology, stellar properties, and SED modeling. More thorough analyses including numerical modeling of disk physical properties will be reported in future papers. Followup observations of these newly seen young disk systems around roughly solar type stars may help elucidate the dynamical processes at work at ages at which terrestrial planets may be forming.   Complementary visible-light imaging of the disks around HD~30447, HD~35841, HD~191089, and HD~141943 is being obtained in \emph{HST} Cycle 21 using the Space Telescope Imaging Spectrograph (STIS) coronagraph  (M. Perrin, PI), and we will be obtaining further infrared observations using the Gemini Planet Imager.  Just as the presence of the exoplanets around HR~8799 and $\beta$~Pic was initially inferred from their infrared excesses due to dust, we recognize a significant likelihood that massive exoplanets exist around these stars as well.

\acknowledgements
This project was made possible by the Mikulski Archive for Space Telescopes (MAST) at STScI. Support was provided by NASA through grants HST-AR-12652.01 (PI: R.~Soummer) and HST-GO-11136.09-A (PI: D. Golimowski) from STScI, which is operated by AURA under NASA contract NAS5-26555. The input images to ALICE processing are from the recalibrated NICMOS data products produced by the Legacy Archive project, ``A Legacy Archive PSF Library And Circumstellar Environments (LAPLACE) Investigation,"  (HST-AR-11279: PI. G. Schneider), and by STScI Director's Discretionary Research funds. The HST data used were originally taken in programs GTO-7226, GO-10177, GO-10527, GO-10849, and GO-10176 (PIs: E. Becklin, G. Schneider, D.C. Hines, S. Metchev, and I. Song). Pueyo was supported in part under contract with the California Institute of Technology (Caltech) funded by NASA through the Sagan Fellowship Program executed by the NASA Exoplanet Science Institute. 


\end{document}